\documentclass[aps,prl,twocolumn,superscriptaddress,showpacs]{revtex4}
\usepackage{epsfig,amsmath,amsfonts}
\newcommand{\nc}{\newcommand}
\nc{\rnc}{\renewcommand}
\nc{\ket}[1]{| #1 \rangle}
\nc{\bra}[1]{\langle #1 |}
\nc{\proj}[1]{\ket{#1}\bra{#1}}
\nc{\braket}[2]{\langle #1| #2 \rangle}
\nc{\hilb}{\mathcal{H}}
\nc{\inprod}[2]{\braket{#1}{#2}}
\def\id{{\mathbf 1}}
\rnc{\vec}[1]{\boldsymbol{#1}}
\begin{document}

\title{Mixed state geometric phases, entangled systems,
and local unitary transformations}

\author{Marie Ericsson}
\affiliation{Department of Quantum Chemistry,
Uppsala University, Box 518, Se-751 20 Sweden}
\author{Arun K. Pati}
\affiliation{Institute of Physics, Bhubaneswar-751005,
Orissa, India}
\author{Erik Sj\"{o}qvist}
\affiliation{Department of Quantum Chemistry,
Uppsala University, Box 518, Se-751 20 Sweden}
\author{Johan Br\"{a}nnlund}
\affiliation{SCFAB, Department of Physics, Stockholm University,
Se-106 91 Stockholm, Sweden}
\author{Daniel. K. L. Oi}
\affiliation{Centre for Quantum Computation, Clarendon Laboratory, University
of Oxford, Parks Road, Oxford OX1 3PU, UK}

\begin{abstract}
  The geometric phase for a pure quantal state undergoing an arbitrary
  evolution is a ``memory'' of the geometry of the path in the projective
  Hilbert space of the system. We find that Uhlmann's geometric phase for a
  mixed quantal state undergoing unitary evolution not only depends on the
  geometry of the path of the system alone but also on a constrained bi-local
  unitary evolution of the purified entangled state. We analyze this in
  general, illustrate it for the qubit case, and propose an experiment to test
  this effect. We also show that the mixed state geometric phase proposed
  recently in the context of interferometry requires uni-local transformations
  and is therefore essentially a property of the system alone.
\end{abstract}
\maketitle

Pancharatnam \cite{pancharatnam56} was first to introduce the concept
of geometric phase in his study of interference of light in distinct
states of polarization. Its quantal counterpart was discovered by
Berry \cite{berry84}, who proved the existence of geometric phases in
cyclic adiabatic evolutions.  This was generalized to the case of
nonadiabatic \cite{aharonov87} and noncyclic \cite{samuel88}
evolutions. The geometric phase was also derived on the basis of
purely kinematic considerations \cite{aitchison92}.  In a general
context, the geometric phase was defined for nonunitary and
non-Schr\"{o}dinger \cite{pati95} evolutions.  Since the geometric
phase for a pure state is a nonintegrable quantity and depends only on
the geometry of the path traced in the projective Hilbert space, it
acts as a memory of a quantum system.

Another important development in this field was initiated by
Uhlmann \cite{uhlmann86} (see also \cite{uhlmann89}), who
introduced a notion of geometric phase for mixed quantal states.
More recently, using ideas of interferometry, another definition
of mixed state phase was introduced in \cite{sjoqvist00a} (see
also \cite{bhandari01}) and experimentally verified in
\cite{du03}. A renewed interest in geometric phases for mixed
states is due to its potential relevance to geometric quantum
computation \cite{pachos99}.

Mixed states naturally arise when we ignore the ancilla subsystem of a
composite object (system+ancilla) that is described by a pure entangled state.
In this Letter we wish to consider the mixed state geometric phases in
\cite{uhlmann86,sjoqvist00a} in terms of such purifications, and to
investigate whether they should be regarded as properties of the system alone
or not. More precisely, we would like to address the following question: do
mixed state geometric phases depend only on the evolution of the system of
interest, or do they also depend on the evolution of the ancilla part with
which the system is entangled? By examining, in detail, the case of mixed
states undergoing local unitary evolutions, we find that the Uhlmann phase
\cite{uhlmann86} indeed contains a memory of the ancilla part, while the mixed
state phase proposed in \cite{sjoqvist00a} does not. In particular, we propose
an experiment to test the Uhlmann phase using a Franson set up
\cite{franson89} with polarization entangled photons \cite{hessmo00,white99}
that would verify this new memory effect. More importantly, we show that the
phase holonomies given in \cite{uhlmann86} and in \cite{sjoqvist00a} are
generically different.

Consider first the unitary path $\eta : t \in [0,\tau ] \mapsto
|\psi_{t} \rangle \langle \psi_{t}|$ of normalized pure state projectors with
$\langle \psi_{0} | \psi_{\tau} \rangle \neq 0$. The geometric phase
associated with $\eta$ is defined as
\begin{equation}
\beta =\arg  \lim_{N \rightarrow \infty}
\Big( \langle \psi_{0} | \psi_{\tau} \rangle \
\langle \psi_{\tau} | \psi_{[(N-1)\tau /N]} \rangle \times \ ... \
\times \langle \psi_{[\tau /N]} | \psi_{0} \rangle \Big)
\label{eq:puregp}
\end{equation}
$\beta$ is a property only of the path $\eta$ as it is independent of the lift
$\eta \longrightarrow \tilde{\eta} : t \in [0,\tau ] \mapsto
|\psi_{t}\rangle$. A parallel lift is defined by requiring that each
$\langle\psi_{[(j+1)\tau/N]}|\psi_{[j\tau /N]}\rangle$ be real and positive
(i.e. $\inprod{\psi}{\dot{\psi}}=0$ when $N\rightarrow\infty$), so that
$\beta$ takes the form
\begin{equation}
\beta=\arg\inprod{\psi_{0}}{\psi_{\tau}}.
\label{eq:pureparallel}
\end{equation}
One may measure $\beta$ in interferometry as a relative phase shift in the
interference pattern characterized by $\nu
e^{i\beta}=\inprod{\psi_{0}}{\psi_{\tau}}$, where $\nu = |\langle
\psi_{0}|\psi_{\tau}\rangle|$ is the visibility~\cite{sjoqvist01}.

To generalize the above to mixed states, consider the path $\zeta : t \in [
0,\tau ] \longrightarrow \rho_{t}$ of density operators
$\rho_{t}$. A standard purification (lift) of $\zeta$ is a path $\tilde{\zeta}
: t \in [ 0,\tau ] \longrightarrow w_{t}$ in the Hilbert space of
Hilbert-Schmidt operators with scalar product $\langle w_{t} , w_{t'} \rangle
= {\text{Tr}} ( w_{t}^{\dagger} w_{t'})$ such that $w_{t} w^{\dagger}_{t} =
\rho_{t}$. Note that $w_{t} = \rho^{1/2}_{t} x_{t}$ is a purification of
$\rho_{t}$ for any unitary $x_{t}$. For a purification where each $| \langle
w_{t} , w_{t'} \rangle |$ is constrained to its maximum
$d[\rho_{t},\rho_{t'}]_{Bures}=\text{Tr}\left[\sqrt{\sqrt{\rho_{t}} \rho_{t'}
\sqrt{\rho_{t}}}\right]$
\cite{uhlmann76}, Uhlmann \cite{uhlmann86} defines the geometric phase
associated with $\zeta$ as
\begin{eqnarray}
\phi_{g} & = & \arg \lim_{N \rightarrow \infty}
\Big( \langle w_{0} , w_{\tau} \rangle \
\langle w_{\tau} , w_{[(N-1)\tau /N]} \rangle
\nonumber \\
 & & \times ... \times \langle w_{[\tau /N]} , w_{0} \rangle \Big).
\end{eqnarray}
The Uhlmann phase $\phi_{g}$ is independent of the purification $\zeta
\longrightarrow \tilde{\zeta}$ as long as it obeys the maximality constraint,
thus $\phi_{g}$ is a property of the path $\zeta$. For pure states $\rho_{t} =
|\psi_{t} \rangle \langle \psi_{t} |$ the constrained purification is
characterized by $\langle w_{t} , w_{t'} \rangle = \langle \psi_{t} |\psi_{t'}
\rangle$ up to an arbitrary phase factor so that $\phi_{g}$ reduces to the
pure state geometric phase $\beta$. A parallel purification is introduced by
requiring that each $w_{[(j+1)\tau /N]}^{\dagger} w_{[j\tau /N]}$ be
hermitian and positive for all $j=0,...,N$. Infinitesimally, this entails that
\begin{equation}
w^{\dagger}_{t} \dot{w}_{t} =  \dot{w}^{\dagger}_{t}w_{t} .
\label{eq:uhlmannpc}
\end{equation}
For such a parallel purification, the geometric phase becomes
\begin{equation}
\phi_{g} = \arg \langle w_{0} , w_{\tau} \rangle ,
\label{eq:uhlmannholonomy}
\end{equation}
which reduces to $\beta$ for pure states.  We show below that $\phi_{g}$ could
be verified in interferometry as a relative phase shift in the interference
pattern characterized by the visibility $|\langle w_{0} , w_{\tau} \rangle |$.

To elucidate the above purification approach, consider the unitary case
$\rho_{0} \longrightarrow \rho_{t} = u_{t} \rho_{0} u_{t}^{\dagger}$. We
introduce a set of eigenvectors $\{ |k\rangle \}$, $k=1,...,N$ with $N$ the
(finite) dimension of Hilbert space, with eigenvalues $\{ \lambda_k \}$ of
$\rho_{0}$ so that
\begin{eqnarray}
w_{0} &=& \rho_{0}^{1/2} =
\sum_{k} \sqrt{ \lambda_k} |k \rangle \langle k|
\nonumber\\
\longrightarrow w_{t} &=& u_{t} \rho_{0}^{1/2} v_{t} =
\sum_{k} \sqrt{ \lambda_k} u_{t}|k \rangle \langle k| v_{t}
\end{eqnarray}
with the unitarity $v_{t} = u_{t}^{\dagger}x_{t}$. With $u_{t}$ and $v_{t}$
related via the parallel transport condition Eq.~(\ref{eq:uhlmannpc}), we
obtain the geometric phase from Eq.~(\ref{eq:uhlmannholonomy}) as
\begin{equation}
\phi_{g} = \arg \sum_{k,l} \sqrt{ \lambda_k \lambda_l}
\langle l |u_{\tau}| k \rangle \langle k |v_{\tau} | l \rangle .
\label{eq:wholonomy}
\end{equation}
The standard purification used by Uhlmann is equivalent to considering a pure
state of the system+ancilla, $w_0\longleftrightarrow
\ket{\Psi_0}\in{\cal H}_s\otimes{\cal H}_a$
evolving under a bi-local operator $u_t\otimes y_t$, in Schmidt form,
\begin{equation}
w_t\longleftrightarrow|\Psi_{t} \rangle =
\sum_{k} \sqrt{ \lambda_k} (u_{t}|k \rangle)
\otimes(y_{t}|k \rangle),
\label{eq:purification}
\end{equation}
where the ancilla unitary $y_t=v_t^T$ (transpose with respect to the
instantaneous eigenbasis of $\rho_t$) obeys the same parallel condition as
before. In this view the geometric phase is given by
\begin{equation}
\phi_{g} = \arg \langle \Psi_{0} | \Psi_{\tau} \rangle .
\label{eq:uhlpurification}
\end{equation}

Let us now consider the case where the composite system undergoes uni-local
unitary transformations so that only the `system' part is affected, i.e.
unitarities of the form $u_{t} \otimes \id$. The purified state now evolves to
\begin{equation}
|\Psi_{t} \rangle =
\sum_{k} \sqrt{ \lambda_k} (u_{\tau}|k  \rangle)
\otimes |k  \rangle
\label{eq:uniphasediff}
\end{equation}
and the phase difference between the initial and final state reads
\begin{equation}
\arg \langle \Psi_{0} | \Psi_{\tau} \rangle =
\arg \sum_{k} \lambda_k \langle k |u_{\tau}|k \rangle
=\arg \text{Tr}[\rho_0 u_{\tau}] .
\label{eq:erik}
\end{equation}
If we require $u_{t}$ to transport each pure state component $|k\rangle$ of
the density matrix in a parallel manner, then
\begin{equation}
\Phi_{g} = \arg \sum_{k} \lambda_k \nu_{k} e^{i\beta_{k}},
\label{eq:interfergp}
\end{equation}
where $\langle k |u_{\tau}|k \rangle = \nu_{k} e^{i\beta_{k}}$ and $\beta_{k}$
is the pure state (noncyclic) geometric phase for $|k\rangle$.
$\Phi_{g}$ is the mixed state geometric phase proposed in \cite{sjoqvist00a}.

It is natural to ask when the two mixed state geometric phases match. To
see this, let us write $u_{t}=\exp(-itH)$ and $v_{t}=\exp(it\tilde{H})$,
$H$ and $\tilde{H}$ being the Hamiltonian of system and ancilla, respectively
(we set $\hbar = 1$). The Hamiltonians $H$ and $\tilde{H}$ are
both assumed to be time-independent. To determine $\tilde{H}$ from the
parallel transport condition Eq.~(\ref{eq:uhlmannpc}), we write $\rho_{0}$ in
its diagonal basis yielding \cite{uhlmann93}
\begin{equation}
\tilde{H} = \sum_{k,l} \frac{2\sqrt{\lambda_{k}\lambda_{l}}}
{\lambda_{k}+\lambda_{l}} |k \rangle \langle l|
\langle k|H|l \rangle .
\label{eq:htilde}
\end{equation}
Now, $v_{t}=\id$ iff $\tilde{H}=0$, which implies that $H=0$ when all
$\lambda_{k}$ are nonvanishing. That is, when all $\lambda_{k}\neq 0$ the two
geometric phases can match only in the trivial case where neither the system
nor ancilla evolve. Thus, in generic cases the two phases are distinct and one
cannot obtain one from the other. However, if $\rho$ is not of full rank,
$\tilde{H}=0$ does not imply $H=0$ in order to match the two geometric phases.
Only in the extreme case of $\rho$ being pure, the two geometric phases are
identical and equal to the standard geometric phase of the system.

It can be seen that Uhlmann's geometric phase is in general a property of a
composite system in a pure entangled state that undergoes a certain bi-local
unitary transformation.  Hence, this geometric phase depends on the history of
the system as well as on the history of its entangled counterpart.  On the
other hand the geometric phase proposed in \cite{sjoqvist00a} requires that
the entangled composite system undergoes a uni-local unitary transformation,
i.e.  the evolution of the ancilla is independent of the evolution of the
system. Thus, this geometric phase is essentially a property of the system
alone; the role of the ancilla is just to make the reduced state of the
system mixed.

It should be noted that the above memory effects are not
equivalent to that of the standard geometric phase acquired by the
purified state, as computed in Ref. \cite{sjoqvist00b}. In fact,
the parallelity condition $\langle
\Psi_{t} | \dot{\Psi}_{t} \rangle = 0$ on the purified state is a
much weaker constraint on the bi-local transformation than
Eq.~(\ref{eq:uhlmannpc}). Indeed, by writing
$|\Psi_{t}\rangle=u_{t}\otimes y_{t}|\Psi_{0}\rangle$ the parallel
transport constitutes a single condition
\begin{equation}
\langle \Psi_{0} | u_{t}^{\dagger} \dot{u}_{t} \otimes \id
|\Psi_{0} \rangle + \langle \Psi_{0} | \id \otimes
y_{t}^{\dagger} \dot{y}_{t} |\Psi_{0} \rangle = 0,
\label{eq:bistandardpc}
\end{equation}
and there are infinitely many $y_{t}$ that fulfill
Eq.~(\ref{eq:bistandardpc}) but not Eq.~(\ref{eq:uhlmannpc}). For uni-local
transformations Eq.~(\ref{eq:bistandardpc}) reduces to
\begin{equation}
\langle \Psi_{0} | u_{t}^{\dagger} \dot{u}_{t} \otimes \id
|\Psi_{0} \rangle = 0,
\label{eq:unistandardpc}
\end{equation}
which is also a weaker condition than that for $\Phi_{g}$. In fact, $\Phi_{g}$
requires that each $\langle k| u_{t}^{\dagger} \dot{u}_{t} |k\rangle$
associated with nonvanishing $\lambda_{k}$ does vanish, while in
Eq.~(\ref{eq:unistandardpc}) only their sum vanishes. Only for $|\Psi \rangle$
being a product state, corresponding to a pure state of the system, the new
memory effects match with the standard geometric phase.

Let us now compute Uhlmann's geometric phase in the noncyclic case for a qubit
(two-level system) undergoing unitary precession.  We assume that the qubit's
Bloch vector initially points in the $z$ direction and has length $r$ so that
$\rho_{0}$ has eigenvalues $\frac{1}{2} (1\pm r)$. Furthermore, assume that
the Hamiltonian of the system is $H= \frac{1}{2} \vec{n}\cdot {\vec{\sigma}} =
\frac{1}{2}(n_{x}\sigma_{x}+n_{z}\sigma_{z})$, $|\vec{n}|^{2} =
n_{x}^{2}+n_{z}^{2}=1$. This determines the Hamiltonian $\tilde{H}$ of the
ancilla via Eq.~(\ref{eq:htilde}) as $\tilde{H} = \frac{1}{2} (\sqrt{1-r^{2}}
n_{x} \sigma_{x} + n_{z} \sigma_{z})$. By introducing the unit vector
$\tilde{\vec{n}} = (\tilde{n}_{x},0,\tilde{n}_{z})$ with the components
$\tilde{n}_{x}= \sqrt{1-r^{2}}n_{x}/\sqrt{1-r^{2} n_{x}^{2}}, \tilde{n}_{z}=
n_{z}/\sqrt{1-r^{2} n_{x}^{2}}$, and the parameter $\tilde{\tau} = \tau
\sqrt{1-r^{2} n_{x}^{2}}$ we obtain the noncyclic Uhlmann phase as
\begin{eqnarray}
\phi_{g} & = & -\arctan \Big( \big( rn_{z} \tan \frac{\tau}{2}  -
r\tilde{n}_{z} \tan \frac{\tilde{\tau}}{2} \big) \Big/
\big( 1 + (n_{z} \tilde{n}_{z}
\nonumber \\
 & & + \sqrt{1-r^{2}} n_{x} \tilde{n}_{x}) \tan \frac{\tau}{2}
\tan \frac{\tilde{\tau}}{2} \big) \Big) .
\label{eq:uhlnoncyclic}
\end{eqnarray}

Let us consider some important special cases. Firstly, the cyclic Uhlmann
phase is obtained by inserting $\tau = 2\pi$ and using $-\tan x = \tan (\pi
-x)$ yielding
\begin{equation}
\phi_{g}=\arctan\left(\frac{rn_{z}}{\sqrt{1-r^{2}n_{x}^{2}}}
\tan\left(\pi\sqrt{1-r^{2}n_{x}^{2}}\right)\right).
\label{eq:uhlcyclic}
\end{equation}
Secondly, in the noncyclic pure state case ($r=1$), we have
$\tilde{\vec{n}}=(0,0,1)$ and $\sqrt{1- n_{x}^{2}} = |n_{z}|$,
which yields
\begin{eqnarray}
\phi_{g}=-\arctan\left(n_{z}\tan(\tau/2) \right) +
\frac{\tau}{2} n_{z} \mod{2\pi} .
\label{eq:noncyclicpure}
\end{eqnarray}
This equals minus one-half of the geodesically closed solid angle of the open
path on the Bloch sphere and is consistent with known expression for the
geometric phase in the case of a pure qubit undergoing noncyclic precession
(see, e.g., Ref.  \cite{klyshko89}). Finally, in the case of the maximally
mixed state ($r=0$), $\tilde{\vec{n}} = \vec{n}$ and $\rho_{0}^{1/2} =
\id/\sqrt{2}$, which yields $w_{0} w_{\tau}^{\dagger} = \rho_{0}$ so
that the geometric phase vanishes, i.e. $\phi_{g} = \arg {\text{Tr}} \rho_{0}
= 0$.

Let us now compare the above results with the mixed state geometric phase in
\cite{sjoqvist00a}. In the diagonal basis $\{ |0\rangle , |1\rangle \}$ of
$\rho_{0}$ we have $\nu_{0} = \nu_{1}$ and $\beta_{0}=-\beta_{1}=-\frac{1}{2}
\Omega$, where $\Omega$ is the geodesically closed solid angle on the Bloch
sphere. For $r\neq 0$, we obtain
$\Phi_{g}=-\arctan\left(r\tan(\Omega/2)\right)$.  These expressions become
identical to those of the Uhlmann approach only for pure states and in the
trivial case $\vec{n}=(0,0,1)$, where neither system nor ancilla
evolve. In the maximally mixed case $\Phi_{g}$ is even indeterminate
as the parallel transport conditions $\langle 0| u_{t}^{\dagger} \dot{u}_{t}|
0\rangle = \langle 1| u_{t}^{\dagger} \dot{u}_{t}| 1\rangle=0$ do not specify
a unique $u_{t}$ for a degenerate density operator, making $\Phi_{g} = \arg
\text{Tr} [ \rho_{0} u_{\tau}] = \arg \text{Tr} [\frac{1}{2} u_{\tau}]$
undefined.

As is clear from Eq.~(\ref{eq:purification}), Uhlmann's geometric phase
retains a memory of the evolution of both system and ancilla due to the
parallelity condition Eq.~(\ref{eq:uhlmannpc}). Using the above purification
scheme $w_{t} \longrightarrow |\Psi_{t} \rangle$, the memory effect associated
with $\phi_{g}$ could be tested experimentally in polarization entangled
two-photon interferometry, as now shall be demonstrated. A detailed
description of the relevant set up shown in Fig.~(1) may be found in
Ref.~\cite{hessmo00}. A photon pair (system and ancilla photon) is
produced in a polarization entangled pure state that takes the Schmidt
form in the horizontal-vertical
$(H-V)$ basis:
\begin{equation}
|\Psi_{0} \rangle  =
\sqrt{\frac{1+r}{2}} |H \rangle \otimes|H \rangle +
\sqrt{\frac{1-r}{2}} |V \rangle\otimes |V \rangle .
\label{eq:puremultiphoton}
\end{equation}
This source is described in Ref.~\cite{white99}, and is used
as input in a Franson interferometer~\cite{franson89}. Note
that $\rho_{0} = {\text{Tr}}_{a} |\Psi_{0} \rangle \langle
\Psi_{0} | = \frac{1}{2} (1+r\sigma_{z})$ in the $H-V$ basis,
and that $|\Psi_{0} \rangle$ is isomorphic to
$w_{0} = \sqrt{\frac{1+r}{2}} |H\rangle \langle H| +
\sqrt{\frac{1-r}{2}} |V\rangle \langle V| $.

\begin{figure}
\epsfig{figure=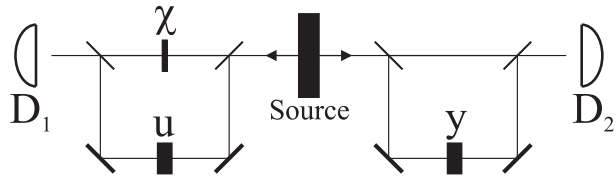,width=0.45\textwidth}
\caption{Two-photon interferometry set up to test the Uhlmann
phase.} \label{fig:figuhl1}
\end{figure}

The two unitary operators $u_{\tau}$ and $y_{\tau}$ are applied to
the two longer arms. Thus, $u_{\tau}$ is applied to the system
photon, say, and $y_{\tau}$ is applied to the ancilla photon. In
one of the shorter arms a $U(1)$ shift $\chi$ is applied. To
observe interference of $|\Psi_{0} \rangle$ and
$|\Psi_{\tau}\rangle = u_{\tau} \otimes v_{\tau} |\Psi_{0}\rangle$
we require that the source produces photon pairs randomly
\cite{franson89}, as is the case with the present type of source.
If the photons arrive in the detector pair simultaneously, they
either both took the shorter path ($\Psi_{0}$) or the longer path
($\Psi_{\tau}$). The state detected in coincidence is the desired
superposition $|\Psi \rangle \sim e^{i\chi} |\Psi_{0}\rangle +
|\Psi_{\tau} \rangle$. The measured coincidence intensity is
proportional to $\langle \Psi | \Psi \rangle \propto 1 + \nu
\cos(\chi - \phi_{g})$, where the visibility is $\nu = |\langle
\Psi_{0}|\Psi_{\tau} \rangle |$. Thus, by varying $\chi$ the
Uhlmann phase $\phi_{g}$ could be tested using this two-photon set
up.

An explicit realization of the operators $u_{\tau}$ and $y_{\tau}$
could be constructed in terms of an appropriate pair of
$\lambda-$plates as follows. The $SU(2)$ part of the effect in the
$H-V$ basis of a $\lambda-$plate making an angle $\theta$ with the
vertical ($V$) axis is given by $u(\alpha ,\theta ) = \exp( -i\frac{\alpha}{2}
\vec{n}_{\theta} \cdot \vec{\sigma)}$ with $\vec{n}_{\theta} = (\sin
[2\theta] ,0, \cos [2\theta] )$. The precession angle $\alpha$ is
proportional to the thickness of the $\lambda-$plate (e.g., $\alpha =
\frac{\pi}{2}$ for a $\frac{\lambda}{4}-$plate). Now, the Uhlmann
phase is obtained by taking $u_{\tau} = u(\alpha ,\theta )$ and
$y_{\tau} = u^{\dagger}(\tilde{\alpha} ,\tilde{\theta} )$, where the
thickness and orientation of the two $\lambda-$plates are related as
$\tilde{\alpha} / \alpha = \sqrt{1-r^{2} \sin^{2}(2\theta)}$ and
$\tan(2\tilde{\theta})=\sqrt{1-r^{2}}\tan(2\theta)$.

In the cyclic case, $\alpha = 2\pi$ and the visibility of the interference
pattern is reduced by the geometric factor
\begin{eqnarray}
\nu&=&\Bigg(\cos^{2}\left(\pi\sqrt{1-r^{2}\sin^{2}(2\theta)}\right)
\nonumber\\
&+&\frac{r^{2}\cos^{2}(2\theta)}{1-r^{2}\sin^{2}(2\theta)}
\sin^{2}\left(\pi\sqrt{1-r^{2}\sin^{2}(2\theta)}\right)\Bigg)^{1/2} .
\end{eqnarray}
Thus, the visibility is reduced by the entanglement of the purified
state.  For maximally mixed states, corresponding to maximally
entangled $\Psi_0$ \cite{strekalov97}, $\tilde{\alpha} = \alpha$ and
$\tilde{\theta} = \theta$ so that $y_{\tau} =
u^{\dagger}_{\tau}(\alpha,\theta )$. Thus one should choose the same
thickness of the two $\lambda-$plates and their half axes being
perpendicular. The scalar product $\langle \Psi_{0} |u_{\tau} \otimes
y_{\tau} |\Psi_{0} \rangle = \langle
\Psi_{0} |u_{\tau} \otimes u^{\dagger}_{\tau} |\Psi_{0} \rangle$ becomes
real-valued and hence $\phi_{g} = 0$. The absence of phase shift
could, e.g., be tested by varying the common angle $\theta$. For
pure states, $\tilde{\alpha} = \alpha \cos 2\theta$ and
$\tilde{\theta}=0\mod{\frac{\pi}{2}}$. This yields the pure
state geometric phase $\phi_{g} = -\frac{1}{2} \Omega$, which
also could be tested in single-photon interferometry
\cite{sjoqvist01}.

The mixed state geometric phase in \cite{sjoqvist00a} could be tested by
canceling the accumulation of local phase changes for each pure state
component in each beam of a single-photon interferometer. Thus, if one of the
beams is exposed to the unitarity $u_{t}$, the other beam should be exposed to
the unitarity $\tilde{u}_{t}$ fulfilling $\langle 0| \tilde{u}_{t}^{\dagger}
\dot{\tilde{u}}_{t} |0\rangle = - \langle 0| u_{t}^{\dagger} \dot{u}_{t}
|0\rangle$ and $\langle 1| \tilde{u}_{t}^{\dagger} \dot{\tilde{u}}_{t}
|1\rangle = - \langle 1| u_{t}^{\dagger} \dot{u}_{t} |1\rangle$
\cite{sjoqvist01}.

To conclude, we have shown that the mixed state geometric phases proposed in
\cite{uhlmann86} and \cite{sjoqvist00a} can be interpreted as two types of
generically distinct phase holonomy effects for entangled systems undergoing
certain local unitary transformations. We have shown that these phase effects
are different from the standard geometric phase of the purified state.  In the
unitary case, the Uhlmann phase depends on the path of the system as well as
on the ancilla undergoing a constrained bi-local unitary operation. This is a
new type of memory effect that is present only for mixed state phase holonomy.
We have proposed an experiment using polarization entangled photons to test
this effect.  The geometric phase in \cite{sjoqvist00a} depends on a certain
uni-local transformation in which the ancilla part does not evolve. Thus, this
geometric phase is essentially a property of the system part alone and is
testable in one-particle interferometry.  We hope that the mixed state phases
would have applications in many areas of physics and future experiments would
test these memory effects.

We would like to thank Artur Ekert for useful suggestions.
The work by E.S. was financed by the Swedish Research Council. D.K.L.O
acknowledges the support
of CESG (UK) and QAIP grant IST-1999-11234.

\end{document}